\documentclass[showpacs,preprintnumbers,amsmath,amssymb]{revtex4}
\usepackage{graphicx}
\usepackage{dcolumn}
\usepackage{bm}

\begin{document}
\title{Unified Treatment of Quantum  Fluctuation Theorem and Jarzynski Equality in terms of microscopic reversibility}

\author{T.Monnai}\email{monnai@suou.waseda.jp}
\affiliation{Department of Applied Physics, Waseda University, 3-4-1 Okubo, Shinjuku-ku, Tokyo 169-8555, Japan \\
}%

\date{\today}
\begin{abstract}
There are two related theorems which hold even in far from equilibrium, namely fluctuation theorem and 
Jarzynski equality.
Fluctuation theorem states the existence of symmetry of fluctuation of entropy production, while Jarzynski
 equality enables us to estimate the free energy change between two states by using irreversible processes.
On the other hand, relationship between these theorems was investigated by Crooks\cite{Crooks} for the classical 
stochastic systems.
In this letter, we derive quantum analogues of fluctuation theorem and Jarzynski equality in terms of 
microscopic reversibility. 
In other words, the quantum analogue of the work by Crooks\cite{Crooks} is presented.
Also, for the quasiclassical Langevin system, microscopically reversible condition is confirmed.
\end{abstract}

\maketitle

\section{INTRODUCTION}
Since the discovery by Evans et al.\cite{Evans}, many types of fluctuation theorem has been presented 
both for deterministic\cite{Evans}\cite{Gallavotti}\cite{JarzynskiFT} and stochastic systems\cite{Crooks}
\cite{KurchanC}\cite{Spohn}\cite{Narayan}\cite{Monnai2}\cite{Crooks2}\cite{Gaspard}.
Experimental verification of fluctuation theorem was also performed\cite{Wang}.
Though the systems, the definition of entropy production, and interpretations (such as distinction of transient
or steady state) are different, fluctuation theorems have a following universal form. :
 \begin{equation}
\frac{P(\Delta S=A)}{P(\Delta S=-A)}\simeq e^{\frac{\Delta S}{k_B}}
\end{equation}
Here, $\Delta S$ is the entropy generated, and $P(\Delta S =A)$ is the probability for $\Delta S =A$.
A few quantum analogues of fluctuation theorem has been proposed \cite{Kurchan}\cite{Monnai}\cite{Jarzynski1}.
In \cite{Kurchan} the probability for the energy change between initial and final states is considered, 
so we need to perform observations twice, while in \cite{Monnai} the spectrum distribution of work operator is considered
 and we need to perform only one observation. In \cite{Jarzynski1}, probability for heat exchanged between two systems is 
considered, and we need to perform observations twice. The features of the two times measurement approach and the 
work operator approach are discussed in \cite{Allahverdyan}. 
   
In 1997, Jarzynski presented nonequilibrium equality for free energy difference(Jarzynski equality)
\cite{Jarzynski} which is a
generalization of the minimum work principle.
Jarzynski equality states that the change of free energy $\Delta F$ is calculated by taking the sample average
 of exponentiated work done on a system $e^{- \beta W}$ along irreversible processes which has a following form:
\begin{equation}
\Delta F=-\frac{1}{\beta}\log\langle e^{-\beta W}\rangle
\end{equation}
Here, $\Delta F$ is the free energy difference, $\beta$ is the inverse temperature, $W$ is the work done on
a system along some irreversible process, and $\langle \rangle$ denotes the sample average.
The remarkable feature of this equality is that the change of equilibrium quantity can be estimated by using 
the irreversible process.
Jarzynski equality was experimentally confirmed by RNA stretching experiment\cite{Liphardt}.
Quantum analogues of Jarzynski equality was proposed by Mukamel\cite{Mukamel} and by Yukawa\cite{Yukawa}.
  Both quantum analogues employ the situation where a system is initially in thermal equilibrium and obeys
the canonical ensemble, and the time evolution is described by the time dependent Hamiltonian $H(t)$.
 In \cite{Mukamel} Jarzyski equality is derived in terms of master equation, while in \cite{Yukawa} the 
density matrix approach is considered. Here we propose a derivation based on the microscopic reversibility(\ref{Detail3}).

For classical stochastic systems, Crooks\cite{Crooks} derived classical 
 fluctuation theorem and Jarzynski equality in terms of microscopic reversibility.
In this letter, we give a unified  derivation of the quantum analogue of fluctuation theorem and Jarzynski equality 
along the thought of  Crooks\cite{Crooks}.
 The ratio of the probability distributions of entropy generated between time forward and time reversal process is considered, and we
 define the  entropy production analogous to that of
 \cite{Crooks}. So, fluctuation theorem  derived here can be considered as a quantum extension of that of \cite{Crooks}.

Our emphasis is on the unified treatment of these theorems in terms of microscopic reversibility. 
This condition is derived for a quasiclassical Langevin system in the appendix. 
\section{Quantum Fluctuation Theorem}
\subsection{the case of thermally isolated system}
In this subsection, we derive a quantum analogue of fluctuation theorem for a system which couples to an external driving, thermally isolated  and
whose Hamiltonian $H(t)$ is invariant under time reversal. 
In order to perform this, we derive the relation (\ref{Detail}).
In \cite{Maes} for the case of time independent system, the relation (\ref{Detail}) was discussed, however, 
this relation is necessary for the derivation of fluctuation theorem here, we show the derivation.  

Let the unitary time evolution operator between $t=0$ and $t=T$ be $\hat{U}_T$.
And let the density matrix at $t=-0$ be $\rho(0)$ and define $\rho(T)\equiv \hat{U}_T \rho(0) \hat{U}_T^\dagger $.
Consider some arbitrary observables $\hat{A}$ and $\hat{B}$  whose eigenstates $|a_n\rangle $ and $|b_m\rangle $ form the normalized 
orthogonal complete set.

Then, we perform the measurements about $\hat{A}$ at $t=+0$ and about $\hat{B}$ at $t=T$.
Firstly, we show the following relation. 
\begin{equation}
\frac{P_F(|a_n\rangle\rightarrow |b_m\rangle )}{P_R(\Theta|b_m\rangle\rightarrow\Theta|a_n\rangle )}=e^{\log\langle a_n|\rho(0)|a_n\rangle -\log\langle b_m|\rho(T)|b_m\rangle } \label{Detail}
\end{equation}  
Here, $P_F(|a_n\rangle \rightarrow |b_m\rangle )$ is the probability that the forward process $\Pi$ whose observed states at time 
$t=0$ and $t=T$ are $|a_n\rangle $ and $|b_m\rangle $ occurs.
$\Theta $ is the antilinear time reversal operator, and $P_R(\Theta|b_m\rangle \rightarrow\Theta|a_n\rangle )$ is the probability that the time
reversed process $\Pi^*$ occurs, namely at $t=0$, we perform the measurement about $\hat{B}$ and find the initial state as $\Theta |b_m\rangle $ according to 
the  ensemble with the time reversed density matrix $\Theta \rho(T)\overleftarrow{\Theta} $ and at $t=T$ perform the measurement about $\hat{A}$ and find the final state as $\Theta |a_n\rangle $.
We note that in \cite{Maes} the average over $P_F(|\alpha_0\rangle\rightarrow |\alpha_T\rangle)$ of the logarithm of the ratio $\frac{P_F(|\alpha_0\rangle\rightarrow |\alpha_T\rangle)}{P_R(\Theta|\alpha_T\rangle\rightarrow\Theta|\alpha_0\rangle)}$
is shown to be equal to the microcanonical entropy change in the case  $|\alpha_0\rangle$ and $|\alpha_T\rangle$ are macro states provided the Hamiltonian is time independent.     

The derivation of (\ref{Detail}) is as follows.
Because the probability we find the initial state as $|a_n\rangle $ is $\langle a_n|\rho(0)|a_n\rangle $,  one has
\begin{eqnarray}
&&\frac{P_F(|a_n\rangle\rightarrow |b_m\rangle )}{P_R(\Theta|b_m\rangle\rightarrow\Theta|a_n\rangle )} \nonumber \\
=&&\frac{\langle a_n|\rho(0)|a_n\rangle |\langle b_m|U_T|a_n\rangle|^2} 
{ \langle b_m|\overleftarrow{\Theta}\Theta \rho(T)\overleftarrow{\Theta} \Theta|b_m\rangle |\langle a_n|\overleftarrow{\Theta}U_{T}^*\Theta|b_m\rangle )|^2} \nonumber \\
=&&e^{\log\langle a_n|\rho(0)|a_n\rangle -\log\langle b_m|\rho(T)|b_m\rangle )}
\end{eqnarray}
Here, as mentioned above, $\hat{U}_T$ is the unitary time evolution operator and its time reversal is $\hat{U}_{T}^*$, $\overleftarrow{\Theta} $ express that anti linear operator $\Theta $ acts to the left, and we used the 
relation for unitary operator $\langle b_m|\hat{U}_T|a_n\rangle=\langle a_n|\overleftarrow{\Theta}\hat{U}_{T}^*\Theta |b_m\rangle $.
This relation is proved as follows :
Because we assumed that the Hamiltonian is time reversal invariant, $\Theta H(t)=H(t) \Theta $, one has 
\begin{eqnarray}
&&\langle a_n|\overleftarrow{\Theta}\hat{U}_T^*\Theta |b_m\rangle \nonumber \\
&&=\lim_{N\rightarrow \infty}\langle a_n|\overleftarrow{\Theta}\hat{U}_1\hat{U}_2..\hat{U}_N \Theta|m\rangle \nonumber \\
&&=\langle b_m|\hat{U}_T|a_n\rangle 
\end{eqnarray}
Here, $\hat{U}_k$ denotes the unitary time evolution operator between time $\frac{k-1}{N}T$ and $\frac{k}{N}T$.

Then one has the quantum analogue of fluctuation theorem for the system without heat bath as shown below. 

\noindent We define a  counterpart of the entropy production $\Delta S$ analogous to the classical entropy production defined in \cite{Crooks}.
\begin{equation}
\Delta S\equiv k_B\log\langle a_n|\rho(0)|a_n\rangle -k_B\log\langle b_m|\rho(T)|b_m\rangle \label{Entropy}
\end{equation}
Of course this definition of entropy production depends on the choice of the observables $\hat{A}$ and $\hat{B}$.
The state is projected into other basis by measurement and different ways of measurements may cause different entropy productions, unlike
in classical system where all observables commute with each other and the value of the entropy production is unique.  
We note that $\Delta S$ is clearly considered as entropy production when these observables diagonalize the density matrices 
$\rho(0)$ and $\rho(T)$ respectively.
For example, when the system is in equilibrium at initial and final time and we choose $\hat{A}$ and $\hat{B}$ as Hamiltonian, this condition is satisfied.
We treat this case in the  section ‡V.
In such cases, $\Delta S$ is considered as entropy production in the sense of \cite{Crooks}, namely
 $\Delta S \equiv -k_B\log\rho_f +k_B\log\rho_i -\frac{Q}{T_B}$.
Here, $\rho_{i/f}$ is probability density of initial/final times, $Q$ is the heat transferred into the system 
from the heat bath, $T_B$ is the temperature of the heat bath.
In comparison to the diagonal representation of von Neumann entropy $S=-\rm{Tr}(\rho\log\rho)=-\Sigma_n \rho_n\log\rho_n$, the quantity $-k_B\log\rho_f+k_B\log\rho_i$ is considered as 
entropy production of the system, and $-\frac{Q}{T_B}$ is the entropy production of the heat bath.
From the assumption that the system is thermally isolated, heat transferred to
 the system $Q$ is $0$.

Then, one obtains a quantum analogue of fluctuation theorem :
\begin{eqnarray}
&&P_F(\Delta S)\equiv \Sigma_{n,m}P_F(|a_n\rangle\rightarrow |b_m\rangle )\delta(\Delta S-k_B(\log\langle a_n|\rho(0)|a_n\rangle -\log\langle b_m|\rho(T)|b_m\rangle ))\nonumber \\
&&=e^{\frac{\Delta S}{k_B}}\Sigma_{n,m}P_R(\Theta|b_m\rangle\rightarrow \Theta |a_n\rangle )\delta(\Delta S+k_B(\log\langle b_m|\rho(T)|b_m\rangle -\log\langle a_n|\rho(0)|a_n\rangle ))\nonumber \\
&&\equiv P_R(-\Delta S)e^{\frac{\Delta S}{k_B}}\label{FT1}.
\end{eqnarray}
Here, $P_F(\Delta S)(P_R(\Delta S))$ is the probability that the counterpart of entropy production along the forward(backward) process is $\Delta S$. 
\subsection{the system with heat bath}
In this subsection, we consider a system coupled to a heat bath whose total Hamiltonian is $H(t)=H_s(t)+H_B+H_i$.
Here, $H_s(t)$, $H_B$, and $H_i$ are the Hamiltonian of system, heat bath and interaction respectively.
Through out this subsection, we choose $\hat{A}$ and $\hat{B}$ as system Hamiltonian $H_s(0)$ and $H_s(T)$ respectively.  And we denote the n'th eigenvalue of
$H_s(t)$ as $E_n(t)$ and corresponding eigenvector as $|n,t\rangle $. 
We restrict ourselves to the system where the microscopic reversible condition is satisfied.
\begin{equation}
\frac{P_F(|n,0\rangle\rightarrow |m,T\rangle )}{P_R(\Theta|m,T\rangle\rightarrow \Theta|n,0\rangle )}=e^{-\beta Q}  \label{Detail3}
\end{equation}
Here, $P_F(|n,0\rangle\rightarrow |m,T\rangle )$ denotes the conditional probability that the system is initially in 
the energy level $|n,0\rangle $ and after some duration time $T$ jumps to the level $|m,T\rangle $.
The distinction of time forward and reversed processes is the same as previous section.
$Q$ is the heat absorbed by the heat bath.
This relation plays an essential role in \cite{Crooks} for the derivation of Fluctuation Theorem. 
Also in quantum system, there are many situations where the microsocopic reversibility does hold.
In appendix, as an physically important system which satisfies this relation, we confirm this microscopic reversibility for the so-called quasiclassical Langevin system \cite{Gardiner1}.
The rest of this section is devoted for the derivation of fluctuation theorem for the system with the heat bath in terms of the relation (\ref{Detail3}).
The total probability $P_F^{Tot}(|n,0\rangle\rightarrow |m,T\rangle )$ that the system is initially in $|n,0\rangle$ and after some duration $T$ found in $|m,T\rangle $ is given as the product of initial state probability and conditional probability.    
\begin{eqnarray}
&&\frac{P_F^{Tot}(|n,0\rangle\rightarrow |m,T\rangle )}{P_R^{Tot}(\Theta|m,T\rangle\rightarrow \Theta|n,0\rangle )}=\frac{\langle n,0|\rho_s(0)|n,0\rangle}{\langle m,T|\overleftarrow{\Theta}\Theta\rho_s(T)\overleftarrow{\Theta}\Theta|m,T\rangle}e^{-\beta Q}=e^{\frac{\Delta S}{k_B}}\nonumber \\
&&\frac{\Delta S}{k_B} \equiv \log\langle n,0|\rho_s(0)|n,0\rangle -\log\langle m,T|\rho_s(T)|m,T\rangle -\beta Q .\label{Detail2}
\end{eqnarray}  
Here, $Q$ is the heat transferred to the system from the heat bath, and defined the reduced density matrix $\rho_s(t)\equiv \rm{Tr_B}(\rho_{Tot})$ as a system density matrix .

Then one has quantum analogue of fluctuation theorem for the system with a heat bath 
\begin{eqnarray}
&&P_F(\Delta S)\equiv \Sigma_{n,m}P_F^{Tot}(|n,0\rangle\rightarrow |m,T\rangle )\delta(\frac{\Delta S}{k_B}-(\log\langle n,0|\rho_s(0)|n,0\rangle -\log\langle m,T|\rho_s(T)|m,T\rangle -\beta Q))\nonumber \\
&&=e^{\frac{\Delta S}{k_B}}\Sigma_{n,m}P_R^{Tot}(\Theta|m,T\rangle\rightarrow \Theta |n,0\rangle )\delta(\frac{\Delta S}{k_B}+(\log\langle m,T|\rho_s(T)|m,T\rangle -\log\langle n,0|\rho_s(0)|n,0\rangle +\beta Q))\nonumber \\
&&\equiv P_R(-\Delta S)e^{\frac{\Delta S}{k_B}}\label{FT2}.
\end{eqnarray}
This relation corresponds to the classical fluctuation theorem derived in \cite{Crooks}.
\section{Quantum Jarzynski equality} 
In this section, we derive quantum Jarzynski equality in terms of microscopic reversibility (\ref{Detail3}).

At first, we set up the framework on which we discuss here.
We assume that the heat bath is large enough and interaction with the system is weak enough 
so that the following conditions do hold.
As the system density matrix, we use the reduced density matrix $\rho_s(t)\equiv \rm{Tr_B}(\rho_{Tot}(t))$.
In order to discuss the free energy change between two equilibrium states, 
we require that the density matrices of the system $\rho_s(0)$ and $\rho_s(T)$ are canonical distribution at the same temperature.
We set observables $\hat{A}$ and $\hat{B}$ as the system Hamiltonian $H_s(0)$ and $H_s(T)$ as previous section and denote the eigenstate of $H_s(0)$ and $H_s(T)$ as $|n,0\rangle $ and $|m,T\rangle $ with the eigenvalues $E_n(0)$ and $E_m(T)$.
Then 
\begin{eqnarray}
&&\langle n,0|\rho_s(0)|n,0\rangle =\frac{1}{Z_1}e^{-\beta \langle n,0|H_s(0)|n,0\rangle} \nonumber \\
&&\langle m,T|\rho_s(T)|m,T\rangle =\frac{1}{Z_2}e^{-\beta \langle m,T|H_s(T)|m,T\rangle}. \nonumber \\
\end{eqnarray}
Here, $Z_1$ and $Z_2$ are the partition functions for $\rho_s(0)$ and $\rho_s(T)$.
Therefore, the relation (\ref{Detail2}) becomes as 
\begin{eqnarray}
&&\frac{P_F^{Tot}(|n,0\rangle \rightarrow |m,T\rangle )}{P_R^{Tot}(\Theta|m,T\rangle\rightarrow\Theta|n,0\rangle )} \nonumber \\
=&&\frac{Z_2}{Z_1}e^{\beta(\langle m,T|H_s(T)|m,T\rangle-\langle n,0|H_s(0)|n,0\rangle -Q)} \nonumber \\
=&&e^{\beta(E_m(T) -E_n(0)-Q)-\Delta F)}
\end{eqnarray}
Here, $\Delta F \equiv -\frac{1}{\beta}\log\frac{Z_2}{Z_1}$ is the change of Helmholtz free energy.
We note that $W\equiv E_m(T)-E_n(0)-Q$ is considered as work externally done on the system.
So one has 
\begin{equation}
\frac{P_F^{Tot}(|n,0\rangle\rightarrow |m,T\rangle )}{P_R^{Tot}(\Theta|m,T\rangle \rightarrow\Theta|n,0\rangle )}=e^{\beta (W-\Delta F)} .
\end{equation}
Taking average of both sides, one has
\begin{eqnarray}
\langle e^{-\beta (W-\Delta F)}\rangle =&&\Sigma_{n,m}P_F^{Tot}(|n,0\rangle\rightarrow |m,T\rangle) \frac{P_R^{Tot}(\Theta|m,T\rangle\rightarrow\Theta|n,0\rangle )}{P_F^{Tot}(|n,0\rangle\rightarrow |m,T\rangle) }\nonumber \\
=&&1.
\end{eqnarray}
 Here, $\langle..\rangle$ denotes the average over the probability $P_F^{Tot}(|n,0\rangle\rightarrow |m,T\rangle ) $.
Therefore, one has 
\begin{equation}
\Delta F =-\frac{1}{\beta}\log\langle e^{-\beta W}\rangle
\end{equation}
This is the relation that we call the quantum Jarzynski equality.

In summary, we derived a quantum extension of Jarzynski equality  and fluctuation theorem in terms of
microscopic reversibility (\ref{Detail3}). And this relation (\ref{Detail3}) is confirmed for quasiclassical Langevin system.
This unified treatment is the quantum version of Crooks' derivation of fluctuation theorem and Jarzynski equality for classical system.

\begin{acknowledgments}
The author is grateful to  Professors S.Tasaki and H.Hasegawa for fruitful discussions.
This work is supported by JSPS Research Fellowships for Young scientists, a Grant-in-Aid for Scientific Research (C) from JSPS, by a Grant-in-Aid
 for Scientific Research of Priority Areas "Control of Molecules in Intense Laser Fields" and 21st
Century COE Program(Holistic research and Education Center for Physics of Self-Organization Systems)
 both from Ministry of Education, Culture, Sports, Science and Technology of Japan.
\end{acknowledgments}
\appendix
\section{microscopic reversibility for quasiclassical Langevin system}
As an physically important system which satisfy the microscopic reversibility (\ref{Detail3}), we confirm this relation for the so-called quasiclassical Langevin system \cite{Gardiner1}.
This section is devoted for the derivation of relation (\ref{Detail3}) for the system which obeys the quasiclassical Langevin equation.
For the classical Langevin system, microscopic reversibility was derived by Narayan and Dhar \cite{Narayan}.
Here, we utilize their result for quantum Langevin system in semi classical regime. To do this, it is most convenient to use the  quantum noise theory by Gardiner \cite{Gardiner}.
Because a straightforward derivation of Langevin equation is known \cite{Gardiner}, we consider a system which interacts with a harmonic reservoir. Let the total Hamiltonian be 
\begin{equation}
H(t)=\frac{p^2}{2m}+V(q,\lambda(t))+\frac{1}{2}\int d\lambda ((p_\lambda-\kappa_\lambda q)^2+\omega_\lambda^2 q_\lambda^2) 
\end{equation}
Here, $q,p$ are the canonical coordinates of system and $q_\lambda, p_\lambda$ are those of heat bath.
$\lambda(t)$ in the potential term is the control parameter corresponding to the external agent.
Note that for simplicity, we consider the coupling between position and momentum, this is equivalent to the position-position coupling, which might be rather natural form. 
The system is assumed to be initially uncorrelated  to the heat bath and the initial density operator of the heat bath $\rho_B$ is assumed to be canonical.  
($\rho(0)=\rho_s\otimes \rho_B$, $\rho_B=\frac{1}{Z}e^{-\beta \int d\lambda \hbar\omega_\lambda (a_\lambda^\dagger a_\lambda+\frac{1}{2})})$.
This model Hamiltonian is standard and one of the ideal examples which describes the system interacting with the heat bath.
Let $Y(t)$ be an arbitrary system operator in Heisenberg picture. 
Firstly, we define the quantity $\mu(t)$ by 
\begin{equation}
\rm{Tr}_s(Y(t) \rho_s\otimes \rho_B)=\rm{Tr}_s(Y\otimes \mu(t))\rho_B
\end{equation}
The equation of motion for $\mu(t)$ (adjoint equation) is given as 
\begin{equation}
\dot{\mu}(t)=A_0\mu(t)+\alpha(t)A_1\mu(t)
\end{equation}
where $A_0\mu(t)\equiv \frac{i}{\hbar}[H_s,\mu(t)]+\frac{i}{2\hbar}[[\gamma \dot{q},\mu(t)]_+,q]$
 , $A_1\mu(t)\equiv \frac{i}{\hbar}[q,\mu(t)]$ and $\alpha(t)\mu(t)\equiv \frac{1}{2}[\xi(t),\mu(t)]_+$.
$\xi(t)\equiv i\int d\lambda \kappa_\lambda\sqrt{\frac{\hbar\omega_\lambda}{2}}(-a_\lambda(0)e^{-i\omega_\lambda t}
+a_\lambda^\dagger (0)e^{i\omega_\lambda t})$ is the Langevin force and here we assume the density of the state is constant $\frac{2\gamma}{\pi}$ so that the adjoint equation be Markov (The Markov approximation).
Then $\xi(t)$ satisfies the Fluctuation Dissipation Theorem,  
\begin{equation}
\langle [\xi(t),\xi(t')]_+\rangle=\frac{2\hbar \gamma}{\pi}\int_0^\infty d\omega \omega \coth(\beta \hbar \omega)\cos\omega(t-t'),
\end{equation}
where the average $\langle...\rangle\equiv \rm{Tr_B}(...\rho_B)$ denotes the average over the bath variables.
This adjoint equation is rewritten in the form of Kramers equation.
Suppose the Wigner function corresponding to $\mu(t)$ is $W(q,p,t)\equiv \int dr \langle q+\frac{r}{2}|\mu(t)|q-\frac{r}{2}\rangle e^{-i\frac{r p}{\hbar}}$.
Then one has 
\begin{equation}
\frac{\partial W}{\partial t}=(-\frac{\partial}{\partial x}\frac{p}{m}+\frac{\partial}{\partial p}(V'(x,\lambda(t))+
\gamma \frac{p}{m}-\alpha(t)))W+(\Sigma_{n=1}^\infty (\frac{i\hbar}{2})^{2n}\frac{\partial^{2n+1}}{\partial x^{2n+1}}V^{2n+1}(x,\lambda(t)))W
\end{equation}
Due to the associative nature of $\alpha(t)$, the first term of this equation is equivalent to the following 
c-number quasiclassical Langevin equation. Note that up to $O(\hbar)$ the second term is negligible and further more, this approximation becomes better in the large friction limit.
\begin{eqnarray}
&&\dot{x}=\frac{p}{m}\nonumber \\ 
&&\dot{p}=-V'(x,\lambda)-\gamma \frac{p}{m}+\tilde{\alpha}(t) \nonumber
\end{eqnarray}
Here $\tilde{\alpha}(t)$ is the c-number stochastic process whose all the moments are equal to that of $\alpha(t)$ respect to the thermal average.
\begin{equation}
\langle \tilde{\alpha}(t)\tilde{\alpha}(t')\rangle = \rm{Tr}_B(\alpha(t)\alpha(t')\rho_B) \hspace{3mm}etc.
\end{equation}
In particular, under the definition of thermal noise operator $\xi(t)$ and the assumption that $\rho_B$ is canonical, we can show that $\tilde{\alpha}(t)$ is Gaussian process.
Thus the mean and variance are sufficient to determine the stochastic feature of $\tilde{\alpha}(t)$.
\begin{eqnarray}
&&\langle \tilde{\alpha}(t)\rangle =0 \nonumber \\
&&\langle \tilde{\alpha}(t)\tilde{\alpha}(t')\rangle =\frac{\gamma \hbar}{\pi}\int_0^\infty d\omega \omega \coth(\beta \hbar \omega )\cos\omega(t-t')\nonumber
\end{eqnarray}
We derived qausiclassical Langevin equation, and then we can use the result of Narayan and Dhar\cite{Narayan} to obtain the relation (\ref{Detail3}).
From the Gaussian nature of $\tilde{\alpha}(t)$, one has the probability functional $P_{F/R}[\tilde{\alpha}(t)]$ for each realization of $\tilde{\alpha}(t)$
\begin{eqnarray}
P_{F/R}[\tilde{\alpha}(t)]&&=C e^{-\frac{1}{2}\int_0^t ds\int_0^t ds' \tilde{\alpha}(s)(\frac{\gamma \hbar}{\pi}\int_0^\infty d\omega \omega \coth(\beta \hbar \omega)\cos\omega(s-s'))^{-1}\tilde{\alpha}(s')} \nonumber \\
&&=Ce^{-\frac{1}{2}\int_0^t ds\int ds' \tilde{\alpha}(s)(\frac{\beta}{2\gamma}\delta(s-s')+O(\hbar^2))\tilde{\alpha(s')}} \nonumber
\end{eqnarray}
$A(s,s')^{-1}$ denote the inverse of $A(s,s')$ as a kernel. 
Here the distinction of time forward and reverse $F/R$ means that the time reversal of the control parameter $\lambda(t)$ is also considered.
As mentioned above, we consider up to $O(\hbar)$ (semi-classical regime) and thus the $O(\hbar^2)$ term is omitted.
Then after the same discussion of \cite{Narayan}, one has the following relation.
\begin{equation}
\frac{P_F(q_f,p_f|q_i,p_i)}{P_R(q_i,-p_i|q_f,-p_f)}=e^{-\beta(H_s(q_f,p_f,\lambda(t))-H_s(q_i,p_i,\lambda(0))-W)}\label{ND}
\end{equation}
Here, $P_{F/R}(q_f,p_f|q_i,p_i)$ is the probability that the initial and final states for the quasiclassical Langevin equation are $(q_i,p_i)$ and $(q_f,p_f)$ respectively.
$H_s(q,p,\lambda(t))=\frac{p^2}{2m}+V(q,\lambda(t))$ is the system Hamiltonian at time $t$ and $W=\int_0^t ds\frac{\partial}{\partial \lambda}V(q,\lambda(s))\dot{\lambda}(s)$ is the work externally done through the control parameter $\lambda(t)$.
Then we choose the initial and final points of the phase space according to the initial and final Wigner functions.
And the probability that the initial and final states are $|n,0\rangle $ and $|m,T\rangle $ is calculated as 
\begin{equation}
P_F(|n,0\rangle \rightarrow |m,T\rangle )=\langle\int dq_i dp_i W_n(q_i,p_i,0)\int dq_f dp_f W_m(q_f,p_f,T) \delta (q_f-q(t,q_i,p_i)) \delta (p_f-p(t,q_i,p_i))\rangle. 
\end{equation}
Here, $q(t,q_i,p_i)$ and $p(t,q_i,p_i)$ are the solution of the quasiclassical Langevin equation with the initial condition $(q(0),p(0))=(q_i,p_i)$ and $\langle\rangle $ denotes the average over $\tilde{\alpha}(t)$. $W_n(q_i,p_i,0)$ and  $W_m(q_i,p_i,T)$ are the initial and final Wigner
 functions. 
In order to utilize the result for the classical system (\ref{ND}), we note that the Wigner function is very localized in semiclassical regime, (in fact for the harmonic potential case is proportional to $e^{-\frac{1}{\hbar}(\frac{p^2}{\sqrt{k m}}+\sqrt{k m} q^2)}$ where $k$ is the strength of the potential).
Therefore one may take a following view.
Only the neighborhood of energy surfaces $H(q_i,p_i,\lambda(0))=E_n(0)$ and $H(q_f,p_f,\lambda(T))=E_m(T)$ does main contribution to the integration above.
The work done on the system $W$ for long enough time duration is nearly constant irrespective of the initial condition $q(0),p(0)$.
And  the exponent of (\ref{ND}) is within this approximation independent from the path $\tilde {\alpha}(t)$.     
Thus one can finally confirm the microscopic reversibility (\ref{Detail3}) for quasiclassical Langevin system.
\begin{equation}
\frac{P_F(|n,0\rangle \rightarrow |m,T\rangle )}{P_R(\Theta|m,T\rangle \rightarrow \Theta|n,0\rangle )}=e^{-\beta(E_m(T)-E_n(0)-W +O(\hbar))}\simeq e^{-\beta Q}
\end{equation}
Here the first equality results from the above view point and the $O(\hbar)$ deviation is caused by quasiclassical approximation due to the symmetrization procedure such as $qp\rightarrow \frac{1}{2}(qp+pq)$ and this result is enough for the purpose of derivation of (\ref{Detail3}).

\end{document}